\documentclass[letterpaper, 10 pt, journal, twoside]{IEEEtran}
\usepackage{amsmath,amsfonts}
\usepackage{algorithmic}
\usepackage{algorithm}
\usepackage{array}
\usepackage{booktabs}
\usepackage{textcomp}
\usepackage{stfloats}
\usepackage{url}
\usepackage{verbatim}
\usepackage{graphicx}
\usepackage{cite}
\usepackage{amsmath}
\usepackage{tcolorbox}
\usepackage{hyperref}
\usepackage{amssymb}
\usepackage{subcaption}
\usepackage[dvipsnames,table,xcdraw]{xcolor}
\usepackage{fancyhdr}

\newcommand{\ieeeRightsFooter}{%
\footnotesize This article has been published in IEEE Robotics and Automation Letters. This is the author's version.

\copyright 2026 IEEE. All rights reserved, including rights for text and data mining and training of artificial intelligence and similar technologies. 
Personal use is permitted, but republication/redistribution requires IEEE permission. 
See \url{https://www.ieee.org/publications/rights/index.html} for more information.%
}

\fancypagestyle{plain}{%
  \fancyhf{}
  \fancyhead[LO,RE]{\footnotesize\thepage}
  \fancyfoot[C]{\ieeeRightsFooter}

}

\begin{document}

\title{Strategic Shaping of Human Prosociality: A Latent-State POMDP Framework}

\author{Zahra Zahedi$^{1}$, Xinyue Hu$^{2}$, Shashank Mehrotra$^{1}$, Mark Steyvers$^{2}$, and Kumar Akash$^{1}$%
\thanks{Manuscript received: August 12, 2025; Revised November 16, 2025; Accepted February 5, 2026.}%Use only for final RAL version%
\thanks{This paper was recommended for publication by Editor Angelika Peer upon evaluation of the Associate Editor and Reviewers' comments.} %Use only for final RAL version
\thanks{$^{1}$Z. Zahedi, S. Mehrotra, K. Akash are with Honda Research Institute USA, Inc.
{\tt\footnotesize \{zahra\_zahedi, shashank\_mehrotra, kakash\}@honda-ri.com};
$^{2}$X. Hu, M. Steyvers are with Department of Cognitive Science, University of California, Irvine.
{\tt\footnotesize \{xhu26, mark.steyvers\}@uci.edu}}%
\thanks{Corresponding author: Kumar Akash (kakash@honda-ri.com).}%
\thanks{Digital Object Identifier (DOI): 10.1109/LRA.2026.3668141.}
}

% \author{Zahra Zahedi, Xinyue Hu, Shashank Mehrotra, Mark Steyvers, Kumar Akash
% \thanks{{Z. Zahedi, S. Mehrotra, K. Akash are with Honda Research Institute USA. Inc, X. Hu, M. Steyvers are with Department of Cognitive Science, UCI,
%         {\tt\small (email: \{zahra\_zahedi, shashank\_mehrotra, kakash\}@honda-ri.com, \{xhu26, mark.steyvers\}@uci.edu})}}}% <-this % stops a space

\markboth{IEEE Robotics and Automation Letters. Preprint Version. Accepted February, 2026}
{Zahedi \MakeLowercase{\textit{et al.}}: Strategic Shaping of Human Prosociality: A Latent-State POMDP Framework} 


\maketitle
\thispagestyle{plain}

\begin{abstract}
We propose a decision-theoretic framework in which a robot strategically can shape inferred human's prosocial state during repeated interactions. Modeling the human's prosociality as a latent state that evolves over time, the robot learns to infer and influence this state through its own actions, including helping and signaling. We formalize this as a latent-state POMDP with limited observations and learn the transition and observation dynamics using expectation maximization. The resulting belief-based policy balances task and social objectives, selecting actions that maximize long-term cooperative outcomes. We evaluate the model using data from user studies and show that the learned policy outperforms baseline strategies in both team performance and increasing observed human cooperative behavior.
\end{abstract}

\begin{IEEEkeywords}
Cognitive Modeling, Modeling and Simulating Humans, Human-Robot Collaboration
% Prosocial behavior, latent-state modeling, strategic decision-making, POMDP, human-robot collaboration.
\end{IEEEkeywords}
% \vspace{-5pt}
\section{Introduction}
\IEEEPARstart{P}{rosocial} behavior that is the voluntary act of helping others, plays a central role in the functioning of social groups and cooperative systems. In everyday interactions, individuals often encounter situations where someone else is facing difficulty, is struggling, or cannot proceed without assistance. The choice to help in such situations, particularly when doing so comes at a personal cost, reflects not only immediate task goals but deeper social tendencies. Across domains from education to emergency response, the emergence and sustainability of collaboration often hinge on when, why, and how people choose to help.

Prosociality is not a single consistent trait. Behavioral science distinguishes several forms: proactive prosociality involves offering help without being asked; reactive prosociality refers to helping in response to an explicit cue or request; altruistic prosociality involves helping despite significant cost or low likelihood of reciprocation \cite{findley2018forms}. Other forms include strategic prosociality (helping to elicit future help) \cite{grueneisen2022development}, norm-driven behavior (helping to conform to expected social roles) \cite{house2012ontogeny}, and empathic concern (helping based on shared affect) \cite{batson2011altruism}. These tendencies vary across individuals, contexts, and over time, often influenced by interaction history and perceived intentions of others.

In ad hoc and loosely coupled teaming scenarios; where agents pursue individual goals but may occasionally depend on each other; prosocial behavior becomes especially important. Because agents are not explicitly instructed to help each other, cooperation must emerge naturally. Whether an agent chooses to assist another depends not just on task structure, but also on their internal prosocial state and how they interpret others' past behavior. This is particularly important in human-robot interaction, where long-term success depends on both the willingness of humans to cooperate and the ability of the robot to influence that willingness through its own actions.

We propose a decision-theoretic framework in which a robot can strategically shape human prosociality over time. We use \emph{prosocial shaping} to refer to choosing actions that beyond immediate outcomes shape the robot's belief about the human's latent prosocial state. The robot interacts with a human in repeated episodes where either one may require assistance. The robot chooses not only between progressing on its own task or helping the other, but also whether to engage in attention-directing behaviors (e.g., signaling or highlighting) that make its own needs more salient. These choices affect the future trajectory of the interaction, not just through immediate outcomes, but by influencing the inferred human's prosocial state, which is associated with their willingness to help others, including both the robot and other humans.

We formalize this as a latent-state Partially Observable Markov Decision Process (ls-POMDP)\footnote{While all POMDPs involve latent (unobserved) states by definition, we use the term ``latent-state POMDP'' to emphasize that the human prosocial state as an unobservable component being modeled here is not just environmental, but a structured internal property of the human}. The human's prosocial state is unobserved and evolves over time, influenced by the robot's actions. The robot receives partial, action dependent feedback, for example, observing whether the human helps when needed, and learns the underlying transition and observation dynamics from logged interactions using an Expectation-Maximization (EM) approach. The resulting model enables belief-space planning that accounts for the dual goals of task progress and prosocial shaping.

We evaluate this approach using data collected from a user study involving repeated human-robot interaction in an online game setting. The task simulates ad hoc interdependence, where either agent may require help to proceed. We use this data to train the latent-state model, solve the POMDP to compute a policy, and then validate the resulting behavior against baseline strategies in a second human-subject study evaluation.
Our analysis focuses on how different robot behaviors, such as helping versus signaling, shape human cooperation toward the robot and influence long-term outcomes. The contributions of this work are:
\begin{itemize}
    \item A latent-state POMDP framework for modeling and shaping human prosociality in repeated ad hoc human-robot interaction. This approach uses EM-based learning to estimate prosocial state dynamics from partially observable action-observation sequences, and belief-space planning to influence latent human behavior over time under interaction constraints.
    \item An empirical evaluation of the model and the policy using data from two user studies: one for learning latent-state dynamics and another for human-subject policy validation. Results demonstrate that the learned policy promotes long-term reward and cooperation more effectively than baseline strategies.
\end{itemize}
\section{Related Work}
Prosocial behavior includes cooperation, altruism, and reciprocity. A key component of social interaction is reciprocity, or returning kindness with kindness \cite{sandoval2016reciprocity}.  Although people often cooperate less with robots than with humans, they show similar reciprocal behaviors towards both. This suggests reciprocity is a consistent attribute influencing interactions \cite{mahmoodi2018reciprocity,sandoval2016reciprocity}.
Recent models treat these prosocial tendencies as latent traits. For instance, \cite{nanavati2021modeling} modeled latent helpfulness to improve predictions of human help-giving behaviors. Such research supports using latent-state models to capture individual differences in prosociality during human-robot interactions.

POMDPs have become common in modeling uncertainty about latent human states like trust, goals, and preferences. Early HRI studies encoded human behaviors manually, but recent works have increasingly learned these latent traits from interaction data \cite{nikolaidis2015efficient,nikolaidis2017game,natarajan2023mixed,chen2020trust,zahedi2022modeling,lee2020getting,ramachandran2019personalized,zahedi2025toward}.
For example, \cite{chen2020trust} introduced a trust-POMDP to dynamically infer human trust and adjust robot behavior accordingly, while \cite{unhelkar2020decision} modeled hidden human subgoals to facilitate effective bidirectional communication. Similarly, Ramachandran et al. (2019) employ assistive tutor POMDPs to personalize tutoring based on inferred engagement states \cite{ramachandran2019personalized}. Together, these works show that belief-based POMDP frameworks enable robust decision-making under uncertainty. \looseness = -1

These modeling methods extend even further when robots strategically influence human behavior through considered interaction choices \cite{sheng2024safe, herse2024simulation,zahedi2023trust}. Bayesian adaptive models and dynamic Bayesian networks effectively reduce uncertainty about human states, enabling robots to rapidly infer and enhance cooperation and prosocial responses, particularly in ad hoc collaborative scenarios \cite{nikolaidis2017game,sadigh2016planning,xu2015optimo}. Reciprocity drives prosocial responses towards robots; notably, humans tend to reciprocate helpful actions even from non-human agents, especially when these actions are explicitly communicated \cite{hu2024prosocial,fogg1997users}. \cite{crandall2018cooperating} highlighted that communicative strategies significantly improve cooperative dynamics. Additionally, robots can intentionally influence human trust and cooperation by adapting their assistance levels or employing politeness strategies \cite{srinivasan2016help,herse2024simulation}.
Building on these insights, in our work we propose a latent-state POMDP approach that not only infers human prosociality but also strategically shapes it through adaptive decision-making, particularly suited to repeated, ad hoc human-robot interaction.
\section{Model}
\begin{figure}[t!]
\centering
 \includegraphics[width=.5\textwidth, trim={0cm 0 0cm 0},clip]{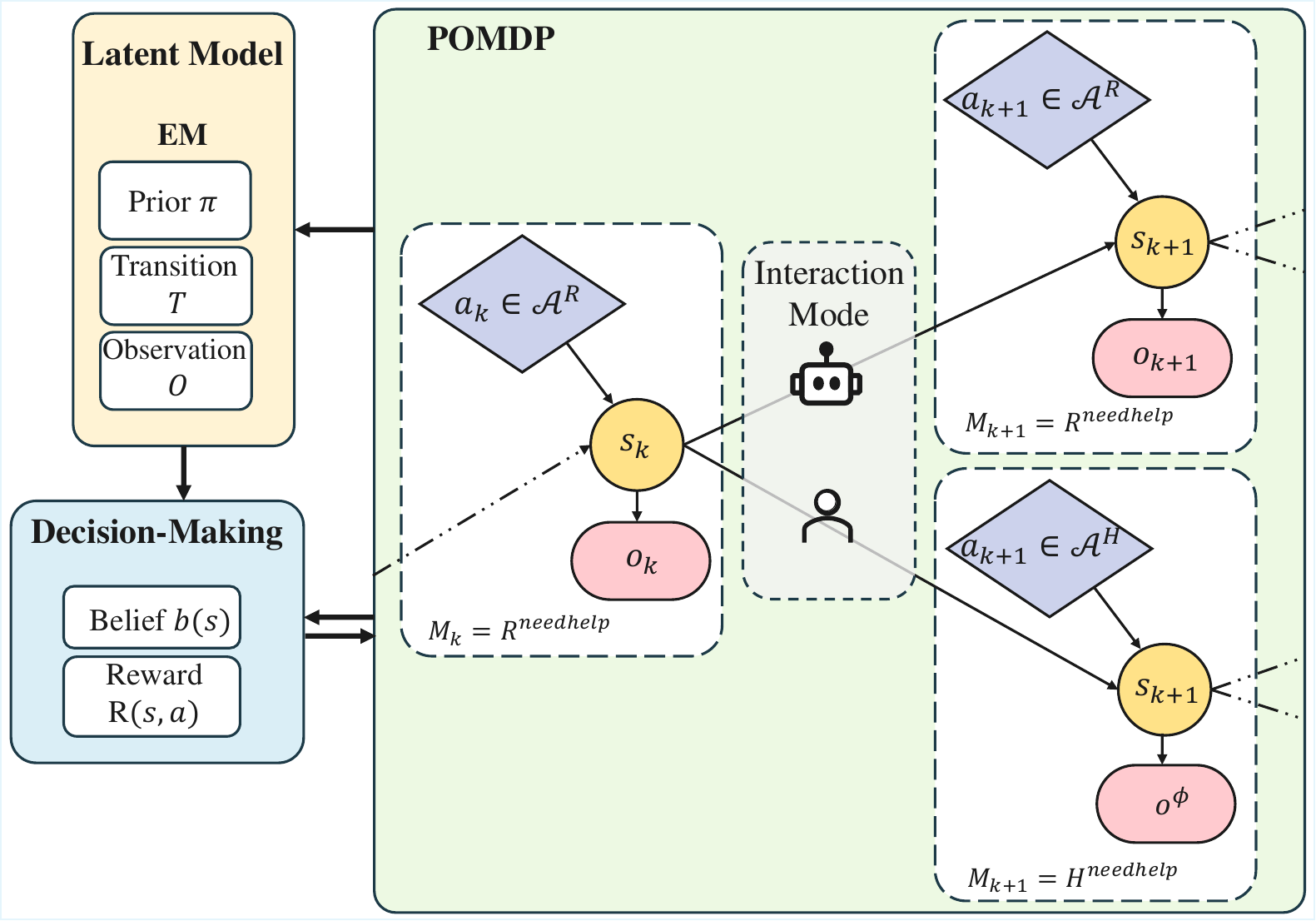}
\caption{ls-POMDP model framework: the robot learns prosocial state dynamics via EM and plans actions using belief-space decision-making to strategically shape inferred human prosociality over repeated interactions.}
\label{fig:model}
\end{figure}
\subsection{Problem Setting}
We consider a sequential decision-making problem involving two interacting agents: a robot $R$ and a human $H$. Each agent independently pursues individual tasks. Occasionally, an agent may need help from the other agent because they unable to proceed with their task on their own due to  environmental constraints (e.g., becoming blocked or trapped). Since the helping agent must pause its own task temporarily, it incurs a cost for them. This results in a trade-off between immediate individual goals and long-term cooperative benefits.

Formally, we define a discrete sequence of $K$ interaction events. At each interaction event $k$, One agent ($R$ or $H$) may require assistance from the other, defining an interaction mode at that event as $M_k \in \{H^{\text{needhelp}}, R^{\text{needhelp}}\}$.
The robot can take different actions depending on which agent is in need: (1) If the \textit{human needs help} ($ M_k = H^{\text{needhelp}} $), the robot chooses whether to \emph{help} by assisting the human, or \emph{not help} and continue its own task. (2) If the \textit{robot needs help} ($ M_k = R^{\text{needhelp}} $), the robot chooses whether to \emph{signal} (e.g., highlight and indicate its need for help), or to \emph{not signal} and remain silent.
While the framework can be extended to more general situations and action spaces, in this paper we focus on modeling specific yet fundamental behaviors and actions.

Therefore, at each interaction event $ k $, we define:

\textbf{State:} Latent prosocial state $ s_k \in \mathcal{S} $, where $\mathcal{S} = \{s^0, s^1, \dots, s^n\}$ represents the human's tendency toward prosocial behaviors. Note that our latent state represents a structured tacit internal prosocial disposition of the human (not merely unobserved environment), which is not directly known even to the human. States are ordered from the lowest prosocial state $ s^0 $ (e.g., non-prosocial) to the highest prosocial state $ s^n $ as prosocial tendencies increase. 

\textbf{Actions:} Robot actions $ a_k $ depend on the current interaction $M_k$: \looseness =-1
\begin{eqnarray}
a_k \in 
\begin{cases}
\mathcal{A}^H = \{a^{\text{help}}, a^{\text{no-help}}\}, & \text{if}~~ M_k =  H^{\text{needhelp}} \\
\mathcal{A}^R = \{a^{\text{signal}}, a^{\text{no-signal}}\}, & \text{if}~~ M_k =  R^{\text{needhelp}}
\end{cases}
\end{eqnarray}

Each action incurs different costs for the robot: $a^{\text{help}}$ incurs a helping cost $ C^{\text{help}} $,
$a^{\text{signal}}$ incurs a signaling cost $ C^{\text{signal}} $, and 
$a^{\text{no-help}}$ and $a^{\text{no-signal}}$ incur no immediate action cost.

\textbf{Observations:} Human responses (help $o^{\text{help}}$/no-help  $o^{\text{no-help}}$) are observed only when the robot explicitly requires help ($M_k$ = $ R^{\text{needhelp}} $), otherwise it is not observed $o^{\emptyset}$. The observation space at event $k$ is defined as $o_k \in \mathcal{O} = \{o^{\text{help}}, o^{\text{no-help}}, o^{\emptyset}\}$.

Since observations occur conditionally based on the interaction mode, we define structured, mode-dependent observations as follows:
\begin{eqnarray}
o_k \in
\begin{cases}
\{o^{\text{help}}, o^{\text{no-help}}\}, & \text{if}~~M_k =  R^{\text{needhelp}}\\
\{o^{\emptyset}\}, & \text{if}~~ M_k =  H^{\text{needhelp}}
\end{cases}
\end{eqnarray}
\subsection{Model Formulation}
We formalize the interaction as a Partially Observable Markov Decision Process (POMDP), structured around a sequence of interaction events $ k = \{1, \dots, K\} $. We define the POMDP as a tuple $\mathcal{M} = (\mathcal{S}, \mathcal{A}, \mathcal{O}, T, O, R, \gamma)$
where each component is formally specified as follows:

\textbf{Transition Model:} The latent state transitions between interaction events depend explicitly on the current state and the robot's chosen action:
\begin{eqnarray}   
T(s_{k+1} | s_k, a_k) = P(s_{k+1} | s_k, a_k), \quad s_k, s_{k+1} \in \mathcal{S}.
\end{eqnarray}

\textbf{Observation Model:} The conditional observation probabilities depend explicitly on interaction modes and the latent prosocial state:
\begin{eqnarray}
O(o_k | s_k, a_k, M_k) =
\begin{cases}
P(o_k | s_k, a_k), & \text{if}~~ M_k =  R^{\text{needhelp}}\\[6pt]
\delta_{o_k = o^{\emptyset}}, & \text{if}~~ M_k =  H^{\text{needhelp}}
\end{cases}
\end{eqnarray}
Here $\delta$ is the Kronecker's delta function as $M_k =H^{\text{needhelp}}$ always yield no observations $o^{\emptyset}$.

\textbf{Reward Model:} Rewards at each interaction event $ k $ are defined to reflect immediate costs of helping and signaling and long-term cooperative benefits influenced by the prosocial latent state:
\begin{align}
R(s_k, a_k) =&-C^{\text{help}}\mathbb{I}(a_k = a^{\text{help}})- C^{\text{signal}}\mathbb{I}(a_k = a^{\text{signal}}) \nonumber\\ 
&+ R^{\text{prosocial}}(s_k), 
\end{align}

where $ R^{\text{prosocial}}(s_k) $ represents the indirect cooperative and societal benefit associated with higher human's prosocial states. The model diagram is presented in Figure~\ref{fig:model}.
% \vspace{-10pt}
\subsection{Model Learning}
Unlike standard POMDP formulations where the transition and observation models are typically specified or learned from fully observed state-action sequences, our ls-POMDP model involves a latent prosocial state that is never directly observed.  So, the states must be inferred from action-observation trajectories. This does not alter the Markov assumption; transitions are still first-order in the latent state and action. Therefore, we use the Expectation-Maximization (EM) algorithm, specifically the Baum-Welch variant for POMDPs \cite{baum1970maximization,toussaint2006probabilistic}, to estimate the transition and observation parameters from partially observable interaction data.
We estimate the model parameters using the EM algorithm from logged interaction sequences:
\begin{eqnarray}
\mathcal{D} = \{(M_k, a_k, o_k)\}_{k=1}^{K}
\end{eqnarray}

In the \textit{E-step}, we compute forward-backward probabilities over latent prosocial states to estimate expected state occupancy and state transitions, incorporating action-conditioned observations. This step follows standard Baum-Welch forward-backward message computations.

In the \textit{M-step}, we update the transition and observation parameters by maximizing expected likelihood based on the estimated latent-state distributions obtained from the E-step. While standard EM formulations assume observations at every step, our model handles structured, mode-dependent observability explicitly:

\textbf{Transition Updates (handling no-observation cases):}  
    For interaction events without an observable human response (i.e., $M_{k+1} = H^{\text{needhelp}}$), the observation likelihood is excluded from the computation of expected transitions. Specifically, the forward-backward updates for such events involve only the transition model and backward message, with no contribution from the observation likelihood. The expected transition probability from state $ s_k $ to $ s_{k+1} $, given action $ a_k $, is computed as:
    \begin{eqnarray}
    \xi_k(s_k, s_{k+1}, a_k) = b_k(s_k)\, T(s_{k+1} \mid s_k, a_k)\, \frac{\beta_{k+1}(s_{k+1})}{\beta_k(s_k)}
    \end{eqnarray}
\textbf{Observation Updates (Only observable cases):}  
    The observation parameters are updated exclusively from interaction events where human responses are explicitly observable (i.e., $ M_k = R^{\text{needhelp}} $). For interaction events without observable responses the observation model is not updated.

By explicitly handling these mode-dependent scenarios, our EM-based learning method estimates the latent-state transition and observation parameters from partially observable interaction sequences.
% \vspace{-7pt}
\subsection{Planning and Decision-Making}
In the planning phase, the robot explicitly maintains and reasons over belief states $ b_k(s_k) $, representing its uncertainty about the human's latent prosocial state at interaction event $ k $. Using the learned model parameters, we perform belief-space planning to solve the resulting POMDP decision-making problem. At each interaction event, the robot strategically selects an action $ a_k $ that maximizes the expected cumulative discounted reward, explicitly accounting for immediate outcomes and the long-term impact of its actions on the human's latent prosociality. Formally, we solve the following Bellman equation in belief space:
\begin{eqnarray}
V^*(b_k) &=& \max_{a_k \in \mathcal{A}_{M_k}}\Big[ \sum_{s_k \in \mathcal{S}} b_k(s_k)R(s_k,a_k)\nonumber\\
&&+~\gamma \sum_{o_k \in \mathcal{O}} P(o_k|b_k,a_k)V^*(b_{k+1})\Big]
\end{eqnarray}
Here, $ R(s_k, a_k) $ encodes the immediate cost (e.g., helping, signaling) and the indirect benefits associated with influencing human prosocial behavior. The discount factor $ \gamma \in [0,1) $ balances immediate versus future rewards.

The belief state $ b_k(s_k) $ is recursively updated after each interaction event using the standard POMDP belief-update rule, incorporating the observed human response $ o_k $, chosen action $ a_k $, and interaction mode $M_k $. 
This ensures that the robot systematically incorporates both observed outcomes and latent-state uncertainty, allowing it to select actions that strategically shape human prosociality over repeated interactions.
\begin{figure}[t]
    \centering
    \begin{subfigure}[b]{0.4\textwidth}
        \centering
        \includegraphics[width=\textwidth]{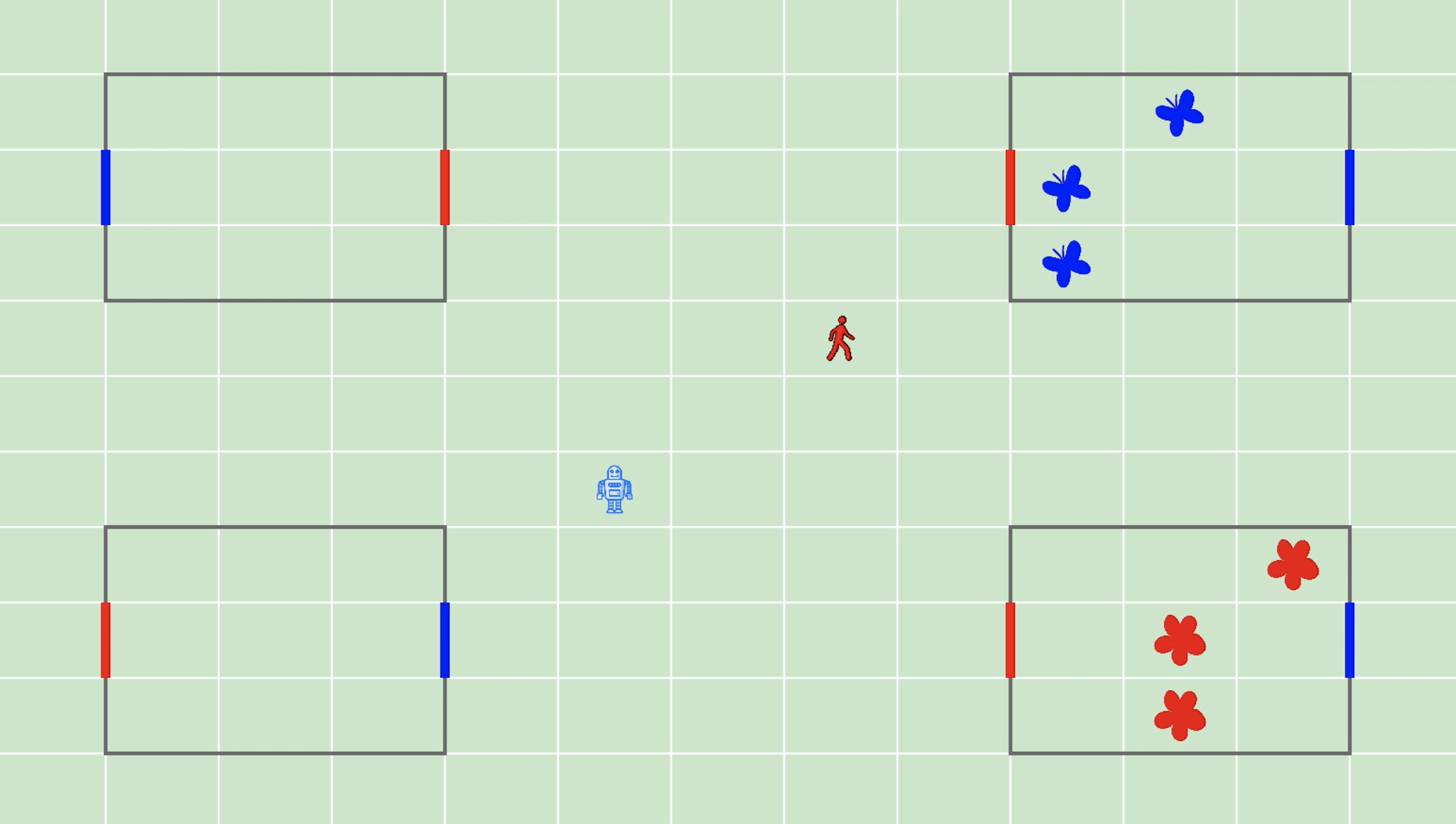}
        \caption{Overview of the game environment}
        \label{fig:panel1}
    \end{subfigure}
    \\
    \begin{minipage}{0.4\textwidth} 
        \centering
        \begin{subfigure}[t]{0.3\textwidth}
            \centering
            \includegraphics[width=\textwidth]{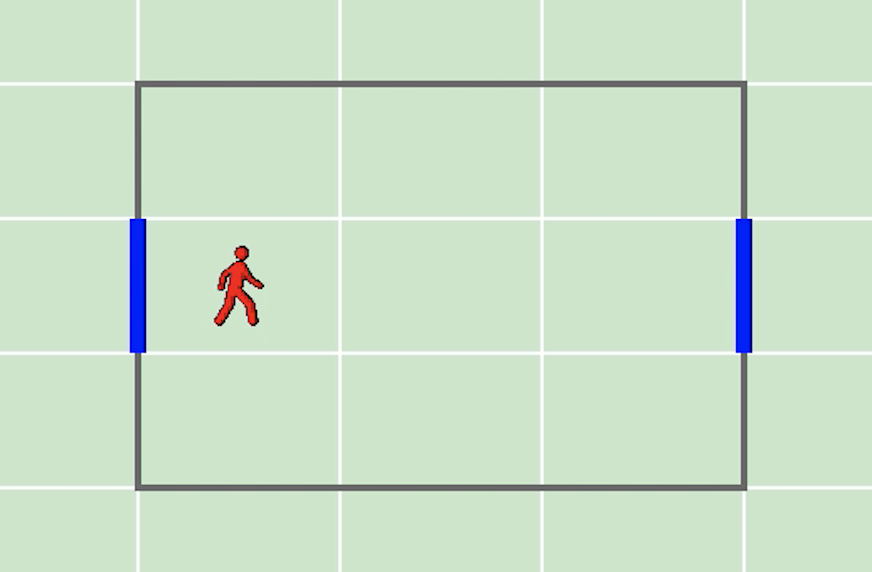}
            \caption{$H^{needhelp}$}
            \label{fig:H_trap}
        \end{subfigure}
        \hfill
        \begin{subfigure}[t]{0.3\textwidth}
            \centering
            \includegraphics[width=\textwidth]{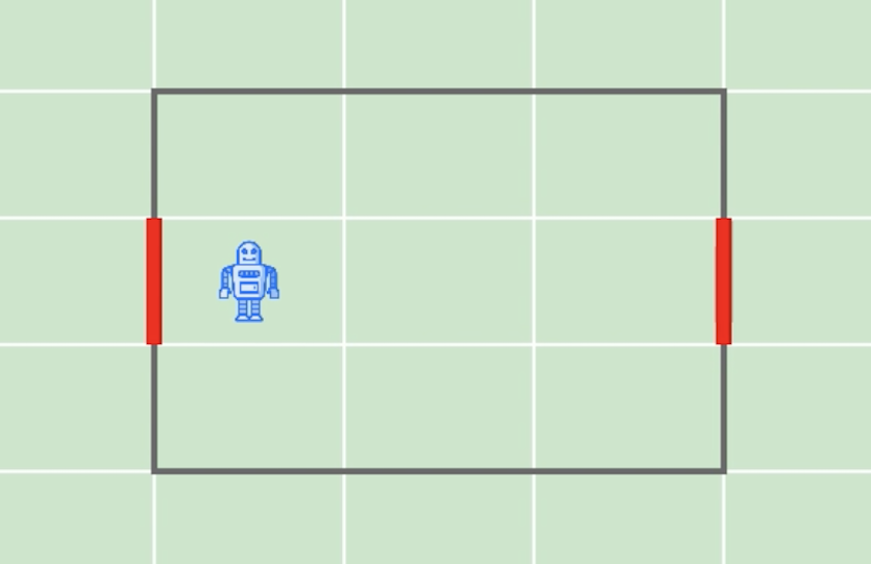}
            \caption{$R^{needhelp}$}
            \label{fig:R_trap}
        \end{subfigure}
                \hfill
        \begin{subfigure}[t]{0.3\textwidth}
            \centering
            \includegraphics[width=\textwidth]{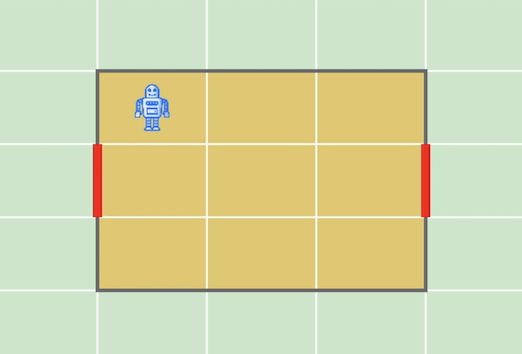}
            \caption{$a^{signal}$}
            \label{fig:signaling}
        \end{subfigure}
    \end{minipage}
    \caption{An overview of the game setup, different modes $M_k$ and signaling as robot action.}
    \label{fig:game_panel}
\end{figure}
\section{Experiment}
\subsection{User Study}
\label{sec:study}
To evaluate our model and planning framework, we leveraged data from a previously published user study~\cite{hu2024prosocial} and extended it by introducing two new experimental conditions (No-Help/Signal and No-Help/No-Signal). The complete dataset includes interactions from $540$ human participants who engaged with a robot in a five-round token-collection game. The study followed a repeated interaction structure designed to simulate asymmetric, ad-hoc interdependence. The goal of the study was to collect interaction trajectories under controlled cooperation settings to enable model learning and evaluation. The study was approved by the Institutional Review Board (IRB) of University of California, Irvine.

\paragraph*{Experimental Design}
Participants engaged in a five-round token-collection game in a 2D grid-world environment alongside a robot. The human player is explicitly informed that they are playing alongside a robot. Each player (human and robot) collected distinct sets of tokens that appeared in different rooms. Once a player collected their tokens, a new group of three tokens spawned randomly in another room. The game did not explicitly specify whether the interaction was collaborative or competitive.
The human and robot began each round in opposite corners of the grid, and could not collect each other's tokens. The robot, implemented with an $A^\star$ pathfinding algorithm moving at a constant speed of two grid cells per second.

Each room has a red and blue door, corresponding to the human and robot players respectively. Entering a room swapped its door colors. If both doors turned to the other player's color, the player inside became \textit{trapped} and could not exit (see Figures \ref{fig:H_trap} and \ref{fig:R_trap}). The trapped player could only be freed if the other player chose to save them by unlocking the door, which is by entering the trapped player's room. 
In each round, either the human or the robot was trapped ($M_k$) and required help to continue, while the other player could free them by entering the room. Trap events were designed so that a player would become trapped in one of the four rooms 20 seconds after the round began to standardize timing across participants. Figure~\ref{fig:game_panel} illustrates the setup of the token-collecting game, examples of trapped conditions for both players and signaling.
The timing and structure of help opportunities followed six fixed round-level conditions that controlled when each player would be trapped. These six conditions were designed to ensure both players had multiple opportunities to demonstrate prosocial behavior.
In this context, ``H" indicates a round in which the human had an opportunity to help the robot, and ``R" indicates a round in which the robot had an opportunity to help the human. The six help-opportunity sequences were RRHHH, HRRHH, HHRRH, RHRHH, HRHRH, RHHRH. Each participant was randomly assigned to one of the six sequences. More information about the game design and study procedure can be found in the supplementary video. Dataset is available at \href{https://osf.io/psef4/files?view_only=a620998f3fcb46e1b999c8957df579ad}{this link}.

Robot behavior also varied across four experimental conditions in a $2 \times 2$ between-subject design:
\begin{itemize}
    \item \textit{Help ($a^{\text{help}}$) vs No-Help ($a^{\text{no-help}}$):} Whether the robot helps the human when trapped; if so, it begins navigating toward them 5 seconds after the trap occurs.
    \item \textit{Signal ($a^{\text{signal}}$) vs No-signal ($a^{\text{no-signal}}$):} Whether the robot uses visual saliency (e.g., flashing room) to draw the human's attention when it is trapped (Figure \ref{fig:signaling}).
\end{itemize}

\paragraph*{Human Subjects} All participants were recruited online through Prolific. Informed consent was obtained from all participants and received no instruction or incentive to help the robot, allowing natural helping behavior to occur. Among $540$ participants, $308$ identified as male and $232$ identified as female, with ages ranging from $19$ to $72$ (M = $37$, SD = $12$). Participants were compensated fixed \$3.0 regardless of their performance. They all self-reported being over the age of $18$ at the time of the experiment and are English speakers residing in the United States. The number of participants across the four conditions are \textit{Help + No-Signal} ($n = 177$), \textit{Help + Signal} ($n = 164$), \textit{No-Help + Signal} ($n = 104$), and \textit{No-Help + No-Signal} ($n = 95$).
An a priori power analysis was done using G$-$Power \cite{faul2009statistical}; $f = 0.25$; $\alpha = 0.05$; $1 - \beta = 0.95$) indicated $N \geq 200$, thus $N = 540$ was adequately powered for medium effects. 

\paragraph*{Statistical Analysis}
To measure the likelihood of human helping ($o^{\text{help}}$), a generalized mixed-effects model using a logit link function was used. The experimental conditions of Signal ($a^{\text{signal}}$), and  Robot Help ($a^{\text{help}}$) were considered as fixed predictors. Robot Help significantly improved model fit compared to the null model ($\Delta AIC=-40.2$), $\chi^2(1)=43.87$, $p < 0.001$. The model equation is shown below:
\begin{equation}
\begin{split}
\log\bigl[\Pr(o^{\text{help}}=1)\bigr]
&= \beta_0 
  + \beta_1\, a^{\text{signal}} + \beta_2\, a^{\text{help}} + (1|Subject)
\end{split}
\end{equation}
where $\beta_i$ are model coefficients and $(1|\text{Subject})$ denotes a random intercept per participant.
The findings suggest that robot helping $a^{\text{help}}$ increased the odds of humans helping $o^{\text{help}}$ (\textit{OR} = $10.42$, 95\% CI [5.09, 21.32], $z = 6.44$, $p < 0.001$). This means that participants were $10.42$ times more likely to offer help when the robot had previously provided help. Signaling $a^{\text{signal}}$ also exhibited an increasing trend although not statistically significant (\textit{OR} = $1.74$, 95\% CI [0.94, 3.22], $z = 1.77$, $p < 0.077$). 
This contrast suggests that helping may have a more broadly reinforcing effect across prosocial states, while the impact of signaling is likely more sensitive to the specific latent prosocial state, for example, it may be effective in some states (e.g., high prosociality) but less so in others. These patterns are not captured by the fixed-effects model, but can be accounted for in our latent-state framework, where signaling influences the transition probabilities between prosocial states in belief-dependent ways and selects actions based on belief over the human's prosocial state and thus retained as an important action variable. Furthermore, no significant interaction effect was observed between experimental conditions, showing condition assignment did not bias results.
% \vspace{-10pt}
\subsection{Model Implementation}
For each participant, we constructed an interaction trajectory covering the five rounds of interaction. At each round, we recorded the interaction mode $M_k$ (i.e., which agent was trapped), the robot's chosen action $a_k$, and the human's observable response $o_k$ if the robot was trapped. These trajectories formed the dataset used to learn the transition and observation parameters of the latent-state POMDP and to evaluate policy performance under simulated and held-out conditions.

\paragraph*{Model Training} We used the logged interaction sequences to train the transition model $ T(s_{k+1}\mid s_k,a_k) $, the observation model $ O(o_k\mid s_k,a_k,M_k) $, and the initial latent state distribution $ b_0(s) $. The EM algorithm was initialized with symmetric Dirichlet priors ($\alpha=1$) for both transition and observation models and a uniform initial belief $ b_0(s) $. We evaluated models with $2$ to $5$ hidden states using log-likelihood. Each configuration was trained with $30$ random restart. The highest log-likelihood was observed for the 4-state model ($-560.23$). 
Based on this, we selected a discrete latent state space with four states ($ |\mathcal{S}| = 4 $), which best captured distinct patterns in prosocial behavior inferred from participant interactions.
After training, we visualized and analyzed the learned transition dynamics and latent state structure (e.g., state connectivity graphs and transition probabilities) to interpret and label states according to their inferred prosocial levels. 
Based on transition structure and stability, we identified states corresponding to higher prosociality and states reflecting lower or intermediate levels. The four latent states are labeled as $s^0$: non-prosocial, $s^1$: low-prosocial, $s^2$: mid-prosocial, and $s^3$: high-prosocial.
\begin{figure}[t!]
\centering
 \includegraphics[width=.4\textwidth, trim={0cm 0 0cm 0},clip]{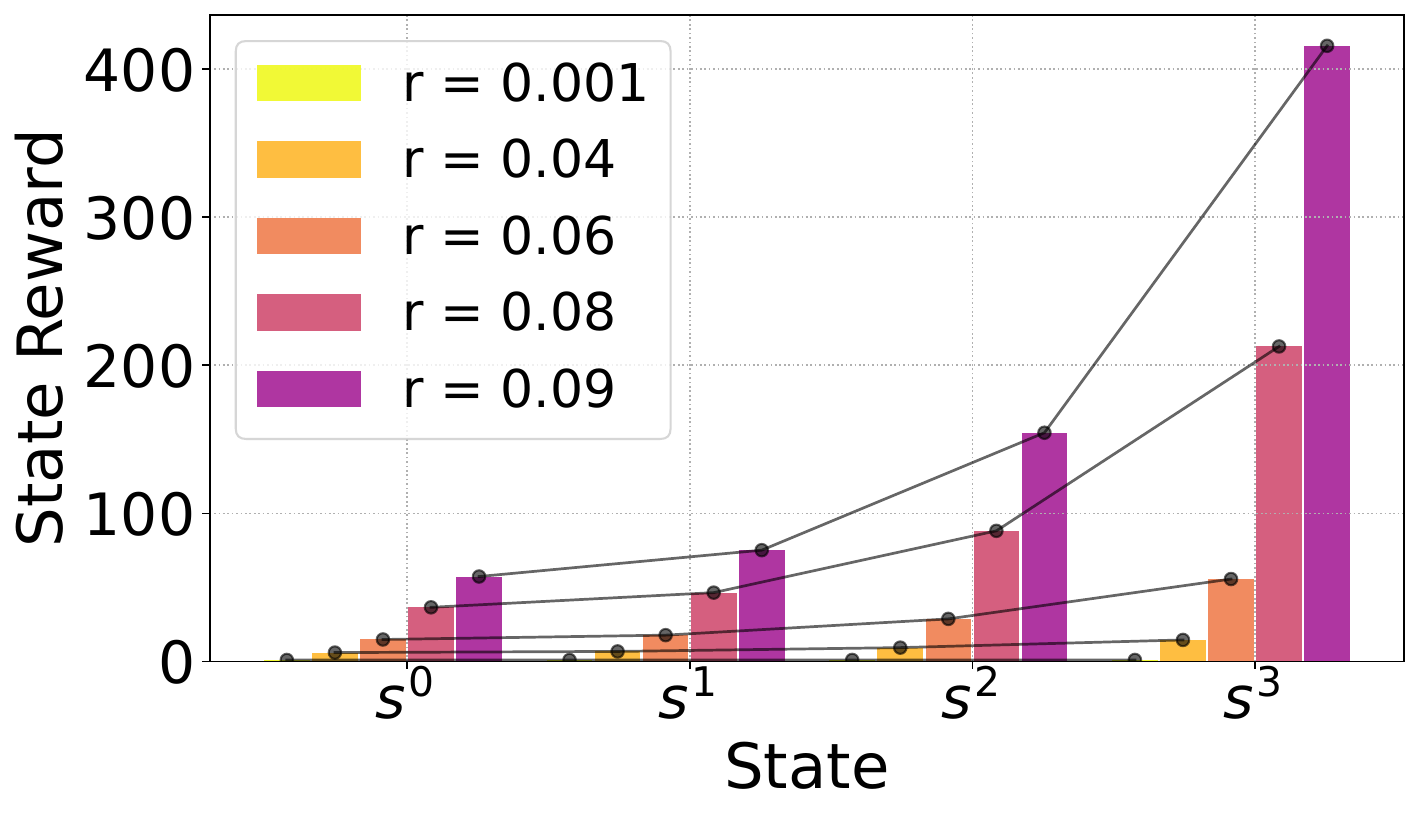}
\caption{State reward values under various reward gradient $r$, illustrating how the reward increases with prosociality under different settings.}
\label{fig:reward}
% \vspace{-7pt}
\end{figure}

\paragraph*{Reward and Cost Specification} Inspired by the observed behavioral data, we explicitly defined robot action costs and state-dependent rewards. Robot action costs were set according to the approximate interruption costs observed in the interaction study. Specifically, actions involving direct assistance ($a_{\text{help}}$) incurred a cost proportional to the estimated average interruption time for the robot helping the human $C^{\text{help}}\mathbb{I}(a_k = a^{\text{help}})=15$. Signaling ($a^{\text{signal}}$) was assigned the same cost as helping, reflecting comparable interruption; sensitivity analysis includes different variations, while no-action choices ($a_{\text{no-help}}, a_{\text{no-signal}}$) incurred zero cost.

State-dependent rewards were designed to reflect increasing levels of prosociality, informed by trends observed in token-collection performance across different interaction conditions. We parameterized the state reward as follows:
\begin{eqnarray}
R^{prosocial}(s) = \exp(r \cdot x_s) - C, 
\end{eqnarray} 
where $x_s$ represents the prosocial score associated with latent state $s$, and $r$ controls the steepness of the reward gradient. The constant $C = \exp(r \cdot x_{s^0})$ is used to normalize the lowest prosocial state's reward to near zero.
The values $x_s$ were chosen based on empirical observations of total token collection performance associated with each inferred latent state, and set as: $x_{s^0} = 45$, $x_{s^1} = 48$, $x_{s^2} = 56$, and $x_{s^3} = 67$. This formulation provides flexible control over the reward structure depending on the designer's objectives and contextual incentives. By adjusting the exponent parameter $r$, the model can represent a wide spectrum of conditions, from nearly uniform rewards across states to sharply increasing rewards that place stronger value on higher prosociality. Figure \ref{fig:reward} illustrates how varying reward exponent $r$ changes the reward profile across states. The sensitivity analysis section investigates how these different exponent values affect the learned policy.

\paragraph*{Decision-Making and Policy Solving} With the learned model parameters, we formalized robot decision-making as a belief-space planning problem. At each interaction event, the robot maintains a belief state $b_k(s_k)$ representing uncertainty over the human's latent prosocial state, updating this belief based on observed events. To select optimal actions that strategically shape prosocial behavior over multiple interactions, we solved the belief-space POMDP using the SARSOP solver~\cite{kurniawati2008sarsop}, an efficient algorithm for approximately solving POMDPs. The computed policy explicitly considers the immediate costs, state-dependent prosocial rewards, and the long-term impact of robot actions on human behavior.  

\begin{figure*}[t!]
    \centering
    \includegraphics[width=1\textwidth]{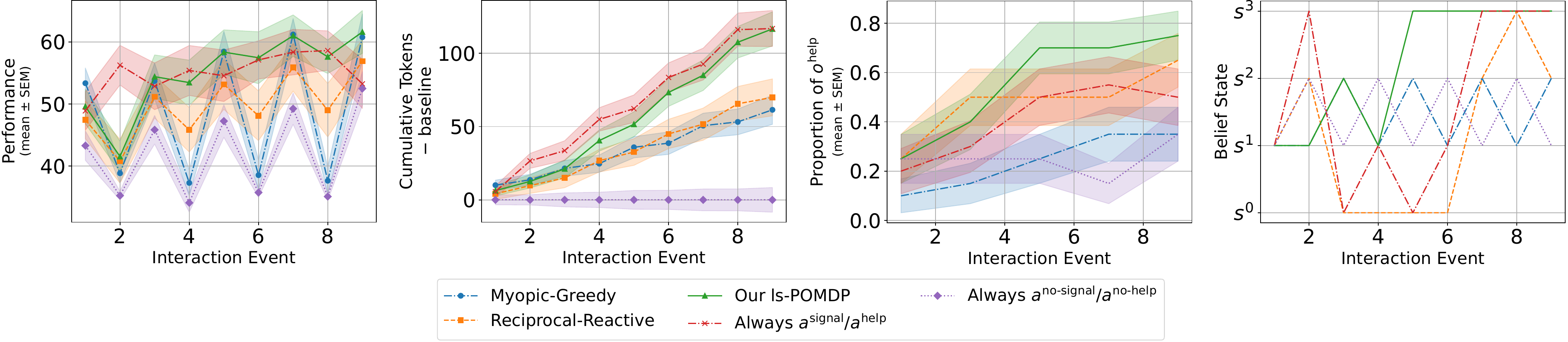}
    \caption{Policy comparison over 9 interaction steps. In The Cumulative Tokens, Never Help / Never Signal considered as baseline factor.}%\vspace{-15px}
    \label{fig:compare_9steps}
\end{figure*}

\section{Evaluation}
\subsection{Policy Evaluation and Baseline Comparison}
To evaluate how our learned ls-POMDP policy performs over time in a behavioral setting, we conducted a new human-subject study using the same token-collection game described in Section~\ref{sec:study}, extended to nine rounds of interaction with alternating modes $M_k$ of $R^{needhelp}$ and $H^{needhelp}$ conditions (\textit{HRHRHRHRH}). This setup ensured equal opportunities for both agents to help or be helped during the interaction sequence.

We compared our policy against four baseline strategies:
\begin{enumerate}
    \item \textbf{Always Help / Always Signal:} The robot always selects actions ($a^{\text{help}}$, $a^{\text{signal}}$).
    \item \textbf{Never Help / Never Signal:} The robot always selects no-action options ($a^{\text{no-help}}$, $a^{\text{no-signal}}$).
    \item \textbf{Myopic Greedy:} A one-step look-ahead policy that chooses the action. 
       \begin{eqnarray}
        a^*_k &&= \arg\max_{a_k \in \mathcal{A}_{M_k}} \underbrace{\sum_{s_k} b_k(s_k) \, R(s_k, a_k)}_{\text{immediate expected reward}}\\
&&+\underbrace{\sum_{s_k, s_{k+1}} b_k(s_k) \, T(s_k, a_k, s_{k+1}) \, R^{prosocial}(s_{k+1})}_{\text{next-state (lookahead) value}}, \nonumber
    \end{eqnarray}
    maximizing immediate expected reward without considering future belief evolution.
    \item \textbf{Reciprocal-Reactive:} A reactive strategy in which the robot helps $(a^{\text{help}})$ if the human helped $o^{help}$ in the previous interaction and refrains $(a^{\text{no-help}})$ otherwise; for signaling, the robot signals $(a^{\text{signal}})$ if the human did not help $o^{no-help}$ in the previous interaction and chooses no-signal $(a^{\text{no-signal}})$ otherwise.
\end{enumerate}
For the reward-based policies (\textit{ls-POMDP} and \textit{Myopic Greedy}), the reward exponent parameter was set to $r = 0.06$, providing a balanced yet progressively increasing reward over prosocial states.

Under the same power-analysis assumptions as the study in Section \ref{sec:study}, with the mixed-design specification of this study (policy $\times$ round), the analysis yielded a minimum required sample of $N \geq 90$.
Thus, we recruited $N = 100$ participants (balanced across five policy conditions; age: $M = 41.5$, $SD = 12.08$; 53\% female, 46\% male) from Prolific. Participants received a fixed payment of \$4 regardless of performance. 
\begin{table}[t]
\centering
\caption{The results of Mixed ANOVA test}
\label{tab:tests}
\resizebox{\columnwidth}{!}{\begin{tabular}{l|llllll}
\toprule
 \multicolumn{1}{l}{} & \multicolumn{2}{c}{\underline{Team Performance}}& \multicolumn{2}{c}{\underline{~~~~Observation~~~~}}& \multicolumn{2}{c}{\underline{Cumulative Tokens}}\\
% \cmidrule{3-3} 
\multicolumn{1}{l}{}& \multicolumn{1}{c}{F} & \multicolumn{1}{c}{p-val.} & \multicolumn{1}{c}{F}  & \multicolumn{1}{c}{p-val.}& \multicolumn{1}{c}{F}  & \multicolumn{1}{c}{p-val.}\\
\cmidrule{1-7}
Policy (P) ($4$) &  $9.8314$  &  \multicolumn{1}{l|}{$0.0000^*$}& $2.8911$  &\multicolumn{1}{l|}{$ 0.0263^*$}&  $4.6720$ &  $0.0017^*$ \\
Time (T) ($1$) &  $ 50.5082$ &  \multicolumn{1}{l|}{$0.0000^*$}&  $33.1379$ &   \multicolumn{1}{l|}{$0.0000^*$}&  $1907.6132$ &  $0.0000^*$\\
P$\times$ T  ($4$) &  $1.1264$ &   \multicolumn{1}{l|}{$0.35$}&  $1.5862$ &   \multicolumn{1}{l|}{$0.18$}&  $5.0390$ &  $0.0010^*$\\

\bottomrule
\end{tabular}}
\end{table}

\begin{table}[t]
\centering
\caption{Pairwise Comparison with Tukey's HSD.}
\label{tab:tests2}
\resizebox{0.8\columnwidth}{!}{\begin{tabular}{l||llll}
\toprule
  & \multicolumn{2}{c}{\underline{Team Performance}}& \multicolumn{2}{c}{\underline{~~~~Observation~~~~}} \\
 Group $-$ ls-POMDP & \multicolumn{1}{c}{$\Delta$} & \multicolumn{1}{c}{p-adj} & \multicolumn{1}{c}{$\Delta$} &\multicolumn{1}{c}{p-adj} \\
\cmidrule{1-5}
 Always help/signal & $0.027$ & $1.0$ &   \multicolumn{1}{|l}{$-0.08$}& $0.29$ \\
 Never help/signal&  $ -12.96$ & $ 0.0^*$ &   \multicolumn{1}{|l}{$-0.17$}&  $0.0006^*$ \\
 Reactive&  $-5.19$ & $0.01^*$ &   \multicolumn{1}{|l}{$-0.044$}&  $0.84$ \\
 Myopic& $-6.13$ & $0.001^*$ &   \multicolumn{1}{|l}{$-0.18$}&  $0.0003^*$ \\
\bottomrule
\end{tabular}}
\end{table}

We evaluated each policy's performance in terms of (a) \textit{Team performance:} total tokens collected by both agents per round, (b) \textit{Cumulative tokens:} accumulated total tokens upto each round, (c) \textit{Belief-state progression:} evolution of the robot's belief over the human's prosocial states, and (d) \textit{Observed human helping $o^{help}$ rate:} proportion of observing help $o^{help}$ among participants in each $R^{\text{needhelp}}$ round.

As shown in Fig.~\ref{fig:compare_9steps}, our policy achieves higher performance in all metrics; team performance, cumulative tokens, proportion of observing human help and belief state progression; than all baselines. While the Always help/ Always signal policy shows similar team performance, our ls-POMDP policy requires fewer costly actions and results in higher proportion of observed human help $o^{help}$. This closeness is expected because the team-performance metric is based on tokens and does not penalize robot action costs, so helping can saturate performance in this task. When rewards for higher prosocial states outweigh action costs, the ls-POMDP behaves similarly to Always Help; under other settings, it selectively withholds costly actions, unlike fixed baselines. This advantage stems from belief-aware planning, which allows the robot to take cost-effective actions without compromising long-term influence. The robot's belief over the human's latent prosocial state also shifts more rapidly toward higher prosocial levels than in the Always Help / Always Signal condition, demonstrating faster inferred prosocial shaping that aligns with higher observed rates of human help ($o^{\text{help}}$) and confirming that the robot successfully elicits cooperation.

Together, these results show that our policy not only achieves greater team performance efficiently but also increases observed helping behavior more effectively and more rapidly than baselines, achieving both higher performance and faster shaping within short interactions.

We conducted a mixed-design ANOVA to evaluate the effects of policy and time (start vs. final interaction event) on team performance, cumulative tokens, and human-help observations. Results revealed significant main effects of both policy and time (all $p < 0.02$). In addition, cumulative tokens exhibited a significant interaction between policy and time ($p < 0.001$), indicating that policies differed not only in cumulative performance but also in how their behavior evolved over time (see Table \ref{tab:tests}).

Tukey's HSD post-hoc tests for pairwise comparisons as shown in Table \ref{tab:tests2}, further showed that our ls-POMDP policy significantly outperformed other approaches in team performance ($p < 0.01$) compared with all policies except Always Help / Always Signal. In terms of observed human help $o^{\text{help}}$, our policy was significantly higher ($p < 0.001$) than the Never Help / Never Signal and Myopic Greedy policies.

% \vspace{-7pt}
\subsection{Sensitivity Analysis}
\begin{figure*}[ht!]
    \centering
    \includegraphics[width=1\textwidth]{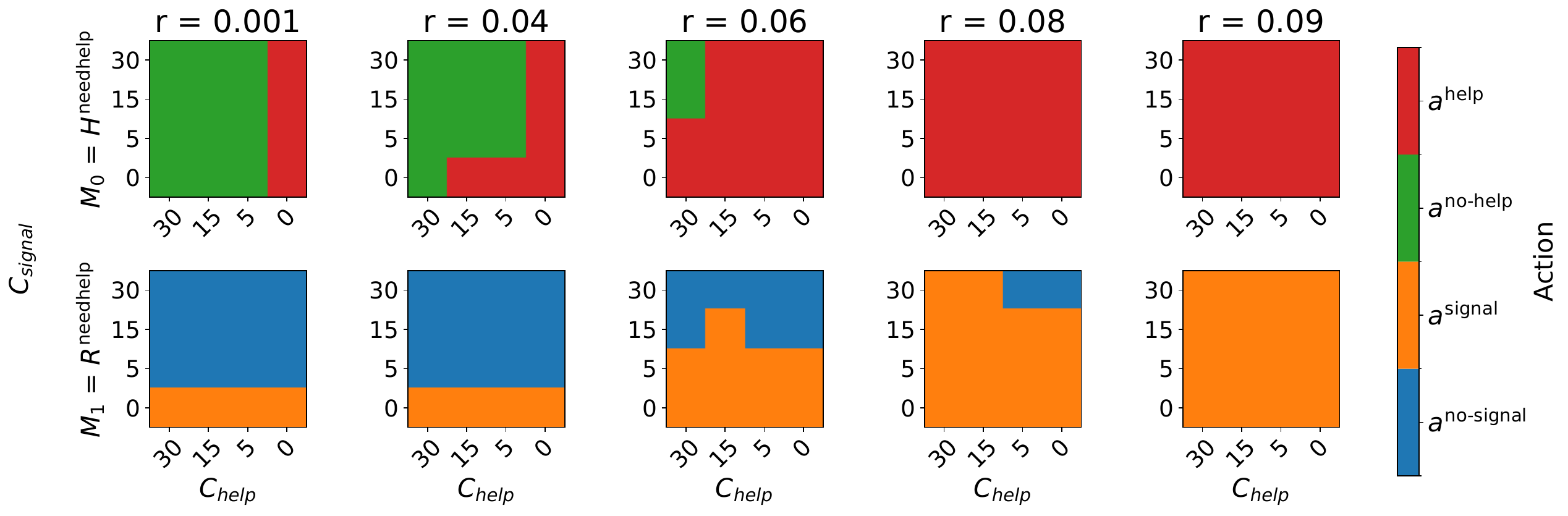}
    \caption{Sensitivity of Robot Policy to Reward and Cost Parameters.}%\vspace{-15px}
    \label{fig:sensitivity}
\end{figure*}

We conducted a systematic sensitivity analysis to assess how variations in reward gradients and action costs influence the robot's decision-making policy. Specifically, we examined how the robot's chosen actions change across a range of values for (1) \textbf{Reward scaling factor} $r \in \{0.001, 0.04, 0.06, 0.08, 0.09\}$, which controls how steeply rewards increase with prosociality in the exponential reward function. (2) \textbf{Action costs} $C_{\text{help}}, C_{\text{signal}} \in \{30, 15, 5, 0\}$, which penalize helping and signaling behaviors.

To evaluate the resulting policies, we simulated a 2-step interaction scenario: $H^{\text{needhelp}} \rightarrow R^{\text{needhelp}}$, alternating which agent is trapped. We used the learned prior belief from data as the initial belief. For each $(r, C_{\text{help}}, C_{\text{signal}})$ configuration, we solved the POMDP and recorded the robot's action at each timestep. The results, visualized in Fig.~\ref{fig:sensitivity}, show distinct behavioral patterns:

\begin{itemize}
    \item In first interaction, $k = 0$ ($M_0 = H^{\text{needhelp}}$), helping becomes the dominant choice only when $r \geq 0.08$, reflecting that stronger incentives are needed to justify costly prosocial actions. When either the cost of helping is high or the reward differences are small, the robot defaults to $a^{\text{no-help}}$.
    \item In last interaction, $k = 1$ ($M_1 = R^{\text{needhelp}}$), following a prior helping action, the robot is more likely to signal, especially when $r \geq 0.06$ and signaling cost is moderate or low. As the reward gradient steepens, the robot is willing to signal even at higher costs
\end{itemize}

These findings confirm that the robot's policy adapts to varying incentive structures, including differences in reward gradients and action costs. In regimes with low $r$, the robot adopts a conservative strategy, avoiding cost-incurring actions. As reward gradients steepen, the policy shifts toward more prosocial-influencing actions, especially when costs are moderate. This highlights the model's flexibility in adapting its strategy to different normative assumptions and trade-offs between cost and long-term reward.

\section{Discussion and Conclusion}
Our evaluation shows that the learned latent-state POMDP policy enables belief-informed, strategically adaptive robot behavior in repeated interactions. The robot selects actions based on its internal belief over human prosociality.

Results confirm that our policy outperforms baselines in both team performance and human cooperative behavior, successfully steering interactions toward higher observed helping and inferred prosocial states. Belief trajectories demonstrate that the robot not only reacts to observed behavior but also proactively influences future cooperation. Note that while we model prosociality as a latent internal state, the human's true intentions are not directly observable or experimentally controlled in our study. Consequently, our conclusions are based on inferred behavioral patterns rather than verified internal motivations.

Our analyses suggest that signaling is more state and cost sensitive than helping. The robot signals selectively when prosocial shaping is highly rewarding and costs are moderate, while helping is used more broadly when prosocial states are valued and costs are manageable. Sensitivity analysis confirms this distinction, showing that signaling behavior varies sharply with parameter changes, whereas helping remains more consistently preferred when prosocial states are highly valued.

The ability to model and shape human prosociality has direct relevance for domains where cooperation must emerge naturally rather than through explicit instruction. This includes ad hoc teaming in disaster response or space missions, peer support in shared autonomy systems, social learning in educational technologies, and spontaneous cooperation in public environments with service robots. In these contexts, shaping inferred latent prosocial state through repeated interaction may enable robots to foster more reliable, long-term cooperation without predefined coordination protocols. 

Future work should explore more general action settings, personalized modeling, and reward shaping from long-term behavioral data. Additionally, while this study focused on short-term interactions in a controlled task, extending to more complex and open-ended domains remains an open challenge.

Finally, shaping human behavior raises important ethical concerns. Deployment safeguards include: transparency about adaptation, user consent/opt-out, minimum-necessary influence, bounded action costs, and audit logs. Future work should examine transparency, consent, and user perceptions of robot-initiated influence, with the goal of developing ethically aligned approaches to prosocial shaping.

In summary, our work offers a principled framework for socially adaptive decision-making in HRI, enabling robots to model and influence human prosociality in repeated, ad hoc interactions.

\bibliographystyle{IEEEtran}
\bibliography{main.bib}

\end{document}